\documentclass[aps,prb,reprint,groupedaddress,footinbib,superscriptaddress, notitlepage]{revtex4-1}
\usepackage{amsmath}
\usepackage{graphicx}
\usepackage{hyperref}
\usepackage{bbold}
\usepackage{color}
\usepackage[normalem]{ulem}

\begin{document}
	\title{Spectral properties of disordered Ising superconductors \\ with singlet and triplet pairing in in-plane magnetic fields}
	\date{\today}
\author{Stefan Ili\'{c}}
\affiliation{Centro de F\'{i}sica de Materiales (CFM-MPC), Centro Mixto CSIC-UPV/EHU, 20018 Donostia-San Sebasti\'{a}n, Spain} 
\affiliation{Univ. Grenoble Alpes, CEA, Grenoble INP, IRIG, PHELIQS, 38000 Grenoble, France}
\author{Julia S. Meyer}
\author{Manuel Houzet}
\affiliation{Univ.~Grenoble  Alpes,  CEA,  IRIG-Pheliqs,  F-38000  Grenoble,  France}
\begin{abstract}
 We study the spectral properties of disordered superconductors with Ising spin-orbit coupling (ISOC) subjected to in-plane magnetic fields. In addition to the conventional singlet pairing, we also consider the recently proposed equal-spin triplet pairing, which couples to the singlet at finite in-plane magnetic fields. While both singlet and triplet order parameters are immune to intravalley scattering, they are significantly affected by intervalley scattering. In the realistic regime of strong ISOC, we find that the properties of the superconductor are well described by a simple formula reminiscent of the well-known Abrikosov-Gor'kov theory, but with a modified self-consistency condition. Our results enable straightforward self-consistent calculation of singlet and triplet order parameters and the density of states of disordered Ising superconductors, which can be particularly useful for interpreting recent tunneling spectroscopy experiments in these systems. We also investigate the high-energy features in the density of states, the so-called mirage gaps, and discuss how they are modified by triplet pairing.  
 \end{abstract}
\maketitle
\section{Introduction \label{Sec1}}

The interplay of superconductivity and spin-dependent fields enables a plethora of exotic phenomena which provide a basis for the emerging fields of superconducting spintronics \cite{linder2015superconducting} and topologically protected quantum computing \cite{alicea2012new}. In the last few years, one of the key platforms for studying such phenomena have been transitional metal dichalcogenide monolayers (TMDs) \cite{lu2015evidence, saito2016superconductivity, xi2016ising, xing2017ising, dvir2018spectroscopy, costanzo2018tunnelling, lu2018full, de2018tuning, sohn2018unusual, li2021printable, cho2022nodal, hamill2021two, kuzmanovic2022tunneling}. These materials are a family of atomically thin superconductors, which host strong intrinsic Ising spin-orbit coupling (ISOC) \cite{zhu2011giant, xiao2012coupled, kormanyos2015k}. ISOC acts as an effective out-of-plane Zeeman field with opposite orientations at the $\pm \mathbf{K}$ corners of the Brillouin zone (valleys). ISOC has remarkable repercussions in superconducting TMDs -- it strongly pins the spins of Cooper pairs to out-of-plane orientation, making TMD superconductors exceptionally robust to in-plane magnetic fields, as experimentally confirmed in various TMDs \cite{lu2015evidence, xi2016ising, saito2016superconductivity, lu2018full, de2018tuning}.

In addition to the conventional singlet pairing, it was recently suggested that TMD superconductors can also support unconventional equal-spin triplet pairing \cite{mockli2019magnetic, mockli2020ising, wickramaratne2020ising}. The dominant singlet order parameter can couple to the triplet order parameter at finite magnetic fields, which significantly affects the high-field behavior of the superconductor. In a recent tunneling spectroscopy experiment \cite{kuzmanovic2022tunneling}, finite triplet pairing was invoked to explain the fact that the superconducting gap was more robust to high magnetic fields than expected from the model with singlet pairing only. 

Another interesting characteristic of Ising superconductors is the appearance of the so-called \lq\lq mirage\rq\rq\ gaps \cite{tang2021magnetic} -- high-energy features in the density of states (DoS). These gaps present a novel experimental signature of unconventional equal-spin triplet correlations enabled by the interplay of ISOC and an in-plane Zeeman field.  Finite triplet pairing provides an additional source of such correlations \cite{kuzmanovic2022tunneling}, which can modify the width of the mirage gaps \cite{patil2023spectral}.

Disorder plays an important role in the properties of Ising superconductors \cite{ilic2017enhancement, mockli2020ising, haim2020signatures}. Both the singlet and equal-spin triplet order parameters are found to be immune to intravalley scattering. On the other hand, both are significantly affected by intervalley scattering. This type of disorder requires a large momentum transfer ($\sim \mathbf{K}$), and therefore it can come from sharp defects in the crystal lattice or edges of the sample. Intervalley scattering acts as an effective spin-flip mechanism which breaks Cooper pairs, namely, electrons scattered from one valley to the other \lq\lq feel\rq\rq\ opposite orientations of the Zeeman field. Moreover, since equal-spin triplet pairing is odd under change of valley, it is fully suppressed by moderate intervalley disorder.

In this work, using the quasiclassical Eilenberger formalism, we establish a theory of spectral properties of disordered Ising superconductors with mixed singlet-triplet pairing subject to in-plane Zeeman fields. For the most part, we will focus on the realistic regime of strong ISOC $\Delta_{so}\gg \Delta, \tau_{iv}^{-1}$, where $\Delta_{so}$ is the energy associated with ISOC, $\Delta$ is the superconducting gap, and $\tau_{iv}^{-1}$ is the intervalley scattering rate. This regime is relevant for most experimentally available Ising superconductors: $\Delta_{so}\gg \Delta$ holds in all superconducting TMDs, while the value of $\tau_{iv}^{-1}$ is sample-dependent, but can be estimated as $\tau_{iv}^{-1}\lesssim \Delta$ in high-quality samples \cite{kuzmanovic2022tunneling}. In this regime, we show that the quasiclassical description of Ising superconductors acquires a particularly simple form, resembling the well known Abrikosov-Gor'kov (AG) theory \cite{abrikosov1960contribution} with a modified self-consistency condition. Our theory provides a straightforward framework for self-consistent calculation of order parameters and the DoS, which can be particularly useful in interpretation of recent experiments such as tunneling spectroscopy measurements \cite{dvir2018spectroscopy, kuzmanovic2022tunneling}. We also investigate the \lq\lq mirage\rq\rq\ gaps at high energies, where our AG-like description does not apply. We discuss how the mirage gaps are modified in the presence of triplet pairing, generalizing the results of Ref.~\onlinecite{patil2023spectral} to the disordered case. 

The paper is organized in the following way. In Sec.~\ref{Sec2}, we introduce the model for disordered Ising superconductors with mixed singlet and triplet pairing. In Sec.~\ref{Sec3} we formulate the quasiclassical Eilenberger equation for this system. In Sec.~\ref{Sec4} we present the main result of our work -- the AG-like equations describing Ising superconductors in the realistic regime of strong ISOC. We use this result to calculate self-consistent order parameters, upper critical field, and the DoS at low energies. Finally, in Sec.~\ref{Sec5}, we investigate the  \lq\lq mirage\rq\rq\ gaps by calculating the DoS at higher energies.

\section{Model \label{Sec2}}
The normal state of an Ising superconductor is described by the following Hamiltonian
\begin{equation}
\mathcal{H}_{\eta}= \xi_\mathbf{q}+\eta \Delta_{so} s_z+ h s_x + V_{0}+V_{iv}.
\label{hamiltonian}
\end{equation}
Here $\xi_{\mathbf{q}}=\mathbf{q}^2/(2m)-\mu$, where $\mathbf{q}$ is the small deviation of the momentum from Dirac points $\pm \mathbf{K}$,  and $\mu$ is the chemical potential. $\Delta_{so}$ is the energy associated with ISOC, $h=g\mu_BB/2$ is related
with the amplitude of the in-plane magnetic field $B$ and the in-plane
$g$-factor ($\mu_B$ is the Bohr magneton), $\eta=\pm1$ is a valley index, and   $s_i$ $(i=x,y,z)$ are Pauli matrices in the spin space. $V_0$ and $V_{imp}$ account for random spin-independent disorder, where $V_0$ describes intravalley scattering with an associated scattering time $\tau_0$, and $V_{iv}$ describes intravalley scattering with a scattering time $\tau_{iv}$. Note that we neglect the contribution of the $\Gamma$ point in the Hamiltonian \eqref{hamiltonian} \cite{xi2016ising,de2018tuning}.

The superconducting order parameter can in general be written as
\begin{equation}
\Psi=\Delta+\mathbf{d} \cdot \boldsymbol{s},
\end{equation}
where $\Delta$ is the singlet order parameter,  $\mathbf{d}=(d_x,d_y,d_z)$ is a vector of the triplet order parameters and $\mathbf{s}$ is a vector of spin Pauli matrices. The general pairing wave function can be written as
\begin{multline}
|\Phi\rangle =\Delta (| \uparrow \downarrow \rangle -|\downarrow \uparrow \rangle )+ d_z (| \uparrow \downarrow \rangle +|\downarrow \uparrow \rangle ) \\
+d_x (| \downarrow \downarrow \rangle -|\uparrow \uparrow \rangle )+ i d_y (| \downarrow \downarrow \rangle +|\uparrow \uparrow \rangle ).
\end{multline}
The component $d_z$ corresponds to opposite-spin triplets, while the components $d_{x,y}$ describe equal-spin triplets. Due to the inversion symmetry breaking,  the $d_z$ triplet may exist in Ising superconductors at zero magnetic field, and its coupling to the singlet is governed by the ratio $\Delta_{so}/\mu$. Since in our model $\mu \gg \Delta_{so}$, the $d_z$ triplet is essentially decoupled from the singlet. Moreover, $d_z$ is suppressed by small intravalley scattering.  On the other hand, applying the Zeeman field enables equal-spin triplets. As shown in Refs.~\onlinecite{mockli2019magnetic, mockli2020ising}, the triplet component that is perpendicular to both the Zeeman field and the ISOC field, $d_y$ in our notation, can couple to the singlet. This singlet-triplet coupling is governed by the ratio $h/\Delta_{so}$, and therefore the presence of the triplet can significantly alter the properties of the superconductor at sufficiently high fields. Moreover, this triplet is immune to intravalley scattering, but since it is odd in the valley index, it is still affected by intervalley scattering. In the following, we will consider an Ising superconductor with dominant singlet pairing and a subdominant $d_y$ triplet with a relative phase of $-\pi/2$~\cite{mockli2019magnetic, mockli2020ising}, parameterized as $d_y=-i\eta\psi $. Other triplets,  $d_x$ and $d_z$, will not be considered, as they do not significantly contribute to field-dependent properties of Ising superconductors.

\section{Quasiclassical equations \label{Sec3}}
If the chemical potential is far above all energy scales relevant for the superconducting properties, $\mu\gg \Delta, \psi, h, \Delta_{so}, \tau_0^{-1}, \tau_{iv}^{-1}$, the Ising superconductor can be described by the quasiclassical Eilenberger equation \cite{tang2021magnetic, mockli2020ising,haim2020signatures}
\begin{equation}
\bigg[(\omega_n+ihs_x)\tau_z+i\eta \Delta_{so} s_z +\Delta\tau_y-\eta\psi s_y\tau_x+\frac1{2\tau_{iv}}g_{\bar\eta},g_\eta\bigg]=0.
\label{eq-Usadel}
\end{equation}
Here, the central object is the quasiclassical Green's function $g_\eta$, which is a  matrix in spin and Nambu spaces, and satisfies the normalization condition $g_\eta^2=1$. Moreover,  $\tau_i$ $(i=x,y,z)$ are Pauli matrices spanning the Nambu space, $\omega_n=2\pi T(n+\frac{1}{2})$ is the Matsubara frequency, and $T$ is the temperature. The effect of intervalley disorder is captured by the last term on the left-hand side of the commutator in Eq.~\eqref{eq-Usadel}. Intravalley disorder has no effect on superconductivity and does not appear in the Eilenberger equation. 

Using Eq.~\eqref{eq-Usadel} and the normalization condition, we find that the quasiclassical Green's function can be parameterized by only two parameters, $x$ and $y$, such that
\begin{widetext}
\begin{equation}
g_\eta=x\left[1+iy\tau_x-i\eta \frac{ \Delta_{so}\frac{\omega_n y-\Delta}{h}-\psi}{\omega_n+\frac{x}{\tau_{iv}}\left(1-\frac{(\omega_n y-\Delta)^2}{h^2}\right)}s_y\tau_y\right]  \left[\tau_z+i\frac{\omega_n y-\Delta}{h}s_x\tau_y\right],
\label{eq-5}
\end{equation}
where $x$ and $y$ are determined by solving the coupled non-linear equations
\begin{equation}
0=\left[\omega_n+\frac{x}{\tau_{iv}}\left(1-\frac{(\omega_n y-\Delta)^2}{h^2}\right)\right]\left[(\omega_ny-\Delta)(\Delta y+\omega_n)+h^2y\right]
- \left(\Delta_{so}\frac{\omega_n y-\Delta}{h}-\psi\right)\left[(\omega_ny-\Delta)\psi-h\Delta_{so}\right],
\label{eq-6}
\end{equation}
\begin{eqnarray}
1&=&x^2\left[1+y^2+\left(\frac{(\omega_ny-\Delta)(\Delta y+\omega_n)+h^2y}{(\omega_ny-\Delta)\psi-h\Delta_{so}}\right)^2\right]\left(1-\frac{(\omega_n y-\Delta)^2}{h^2}\right).
\label{eq-7}
\end{eqnarray}
\end{widetext}

 Using Eq.~\eqref{eq-5}, we can write the coupled self-consistency conditions for the singlet and triplet order parameters as
\begin{align}
\Delta&=\frac{\pi T g_s}{4}\sum_{\eta, \omega_n>0} \text{Tr}[\tau_y g_\eta]= 2\pi T g_s \sum_{\omega_n>0} yx, \nonumber \\
\psi&=\frac{\pi T g_t}{4}\sum_{\eta, \omega_n>0} \text{Tr} [\eta s_y \tau_x g_\eta] \nonumber \\
&= 2 \pi T g_t \sum_{\omega_n>0}\frac{(\omega_ny-\Delta)(\Delta y+\omega_n)+h^2y}{h\Delta_{so}-(\omega_ny-\Delta)\psi}x.
\label{eq-selfcons}
\end{align}
Here $g_{s,t}=\nu_{0} \lambda_{s,t}$, with the normal density of states $\nu_0$ and the coupling constants $\lambda_{s,t}$ in the singlet and triplet channel. Density of states can be obtained after analytical continuation $i\omega_n \to \epsilon+i0^+$, where $\epsilon$ is the energy: 
\begin{equation}
\nu(\epsilon)= \frac{\nu_0}{4}\sum_\eta\text{Re}\text{Tr} [\tau_z g_\eta]_{i\omega_n\to \epsilon+i0^+}= \nu_{0}\text{Re}\left[x\right]_{i\omega_n\to \epsilon+i0^+}.
\end{equation}

 The non-linear system of equations \eqref{eq-6} and \eqref{eq-7} yields multiple solutions, so one needs to select the physical ones. Namely, when solving in Matsubara frequency domain, one should choose real positive solutions ($x,y>0$). When solving in energy domain, we choose the solutions yielding the positive DoS: $\text{Re}[x]\geq 0$. In general, Eqs.~\eqref{eq-6}-\eqref{eq-selfcons} need to be solved simultaneously numerically, which can present a demanding computational problem.  Fortunately,  significant simplifications are possible in the realistic regime where spin-orbit coupling is large, which we discuss in detail in Sec.~\ref{Sec4}. For completeness, in Appendices we present other parameter regimes where simple solutions can be found: in the absence of intervalley disorder for any ISOC strength in Appendix \ref{App1}, and for very strong intervalley disorder  $\tau_{iv}^{-1}\gtrsim \Delta_{so}\gg \Delta$ in Appendix \ref{App3}.

\section{Regime of strong ISOC: $\Delta_{so}\gg \Delta, \tau_{iv}^{-1}$ \label{Sec4}}

In this Section we focus on the realistic regime of strong Ising SOC. First, in Sec.~\ref{Sec4a} we present the simplified AG-like theory for this regime. These results are then used to calculate the self-consistent order parameters (Sec.~\ref{Sec4b}), upper critical field (Sec.~\ref{Sec4c}), and the DoS (Sec.~\ref{Sec4d}).

\subsection{AG-like equation and the self-consistency condition \label{Sec4a}}
In the limit $\Delta_{so}\gg \tau_{iv}^{-1},\Delta$, Eqs.~\eqref{eq-6} and \eqref{eq-7} greatly simplify, and reduce to a single equation for the quantity $u= \Delta_{so} \rho^{-1} y^{-1}$:
\begin{eqnarray}
\frac{\omega_n}{\tilde{\Delta}_{st}}&=&u\left(1-\frac{\tilde{\alpha}}{\tilde{\Delta}_{st}}\frac1{\sqrt{1+u^2}}\right),
\label{eq-AG}
\end{eqnarray}
while $x=u/\sqrt{1+u^2}$, and we introduced $\rho=\sqrt{\Delta_{so}^2+h^2}$. This is the Abrikosov-Gor'kov equation \cite{abrikosov1960contribution}, with the depairing parameter $\tilde{\alpha}$, and a renormalized gap parameter $\tilde{\Delta}_{st}$ accounting for both singlet and triplet pairing:
\begin{equation}
\tilde{\alpha}=\frac{h^2}{\rho^2\tau_{iv}}, \qquad \tilde{\Delta}_{st}=\frac{\Delta_{so}}{\rho}\Delta+\frac{h}{\rho}\psi.
\end{equation}
 AG theory was originally derived to describe superconductors with magnetic impurities, but its validity has  since been extended to many situations where time-reversal symmetry is broken and a pair-breaking mechanism that mixes time-reversed states is present \cite{maki2018gapless}. In our case, time-reversal symmetry is broken by the magnetic field, while intervalley scattering provides the pair-breaking mechanism. In most superconductors described by the AG theory, the depairing parameter is quadratic or linear in the time-reversal symmetry breaking field. In Ising superconductors, the dependence is more complex, and moreover, the renormalized gap parameter $\tilde{\Delta}_{st}$ also depends on the field.    

Another difference compared to the standard AG theory is that the self-consistency condition in Ising superconductors is more complicated. The coupled self-consistency conditions \eqref{eq-selfcons}  at temperatures smaller than the triplet critical temperature, $T<T_{ct}$, simplify to
\begin{align}
0&=\sum_{\omega_n>0}\bigg[\frac{1+\frac{\omega_n^2}{\omega_n^2+\rho^2}\left(\frac{\rho\Delta}{\Delta_{so}\tilde\Delta_{st}}-1\right)}{\sqrt{1+ u^2}}- \frac{\frac{\rho\Delta}{\Delta_{so}}}{\sqrt{\omega_n^2+\Delta_0^2}}\bigg],\nonumber \\
0&=\sum_{\omega_n>0}\bigg[\frac{1+\frac{\omega_n^2}{\omega_n^2+\rho^2}\left(\frac{\rho\psi}{h\tilde\Delta_{st}}-1\right)}{\sqrt{1+ u^2}}-\frac{\frac{\rho\psi}{h}}{\sqrt{\omega_n^2+\psi_0^2}}\bigg].
\label{eq-self}
\end{align}
Here $\Delta_0$ and $\psi_0$ are the singlet and triplet order parameters at $h,\Delta_{so}=0$, at the given temperature $T$, and we assume $\Delta_0>\psi_0$. We used the fact that the coupling constants can be expressed as $g_s^{-1}=2\pi T \sum_{\omega_n} (\omega_n^2+\Delta_0^2)^{-{1}/{2}}$, while at $T<T_{ct}$ we have $g_t^{-1}=2\pi T \sum_{\omega_n} (\omega_n^2+\psi_0^2)^{-1/2}$. On the other hand, at temperatures $T>T_{ct}$, we need to do a substitution $\sum_{\omega_n}1/\sqrt{\omega_n^2+\psi_0^2}\to \sum_{\omega_n}1/{\omega_n}-(2\pi T)^{-1}\ln T_{ct}/T$ in Eq.~\eqref{eq-self} [and also in $F(\psi_0)$ defined  below]. The terms proportional to $\omega_n^2$ in Eq.~\eqref{eq-self} must be kept to ensure convergence at high energies. Combining the two lines of Eq.~\eqref{eq-self}, we can compactly write a single equation for $\tilde{\Delta}_{st}$ as
\begin{equation}
\frac{F(\Delta_0)F(\psi_0)}{\Delta_{so}^2F(\Delta_0)+h^2 F(\psi_0)}=\sum_{\omega_n>0}\frac{1}{\sqrt{1+u^2}}\frac{1}{\omega_n^2+\rho^2},
\label{eq-Deltast}
\end{equation}
 where we defined the function $F$  as
\begin{equation}
F(x)=\sum_{\omega_n>0}\bigg[\frac{1}{\sqrt{1+u^2}}-\frac{\tilde{\Delta}_{st}}{\sqrt{\omega_n^2+x^2}}\bigg].
\end{equation}

If intervalley scattering is strong, $\tau_{iv}^{-1}\gg \Delta_0$, the triplet pairing is fully suppressed, and  we can do the following substitutions in Eq.~\eqref{eq-AG}:
 \begin{equation}
 \tilde\alpha\to \alpha=h^2/(\Delta_{so}^2\tau_{iv}), \qquad \tilde{\Delta}_{st}\to \Delta. 
 \label{eq-17}
 \end{equation}
Moreover, the self-consistency condition in this regime simplifies to 
\begin{equation}
\sum_{\omega_n>0}\bigg[\frac{1}{\sqrt{1+u^2}}-\frac{\Delta}{\sqrt{\omega_n^2+\Delta_0^2}}\bigg]=0,
\label{eq-18}
\end{equation}
and at zero temperature and in the gapped regime ($\alpha<\Delta$) we have $\ln \Delta/\Delta_0=-\alpha\pi/(4\Delta)$. Therefore, in this regime the Ising superconductor is described by the standard AG theory, with the depairing parameter quadratic in field and the standard self-consistency condition.

Eqs.~\eqref{eq-AG} and \eqref{eq-Deltast}  are the main results of this work, allowing for straightforward self-consistent calculations of order parameters for  Ising superconductors in the realistic regime of strong ISOC. Using this result, existing AG solvers can be readily adapted for Ising superconductors with a few changes. The DoS is also readily found using our theory, as detailed in Sec.~\ref{Sec4d}.

\subsection{Self-consistent order parameters \label{Sec4b}}

\begin{figure}
    \includegraphics[width=0.35\textwidth]{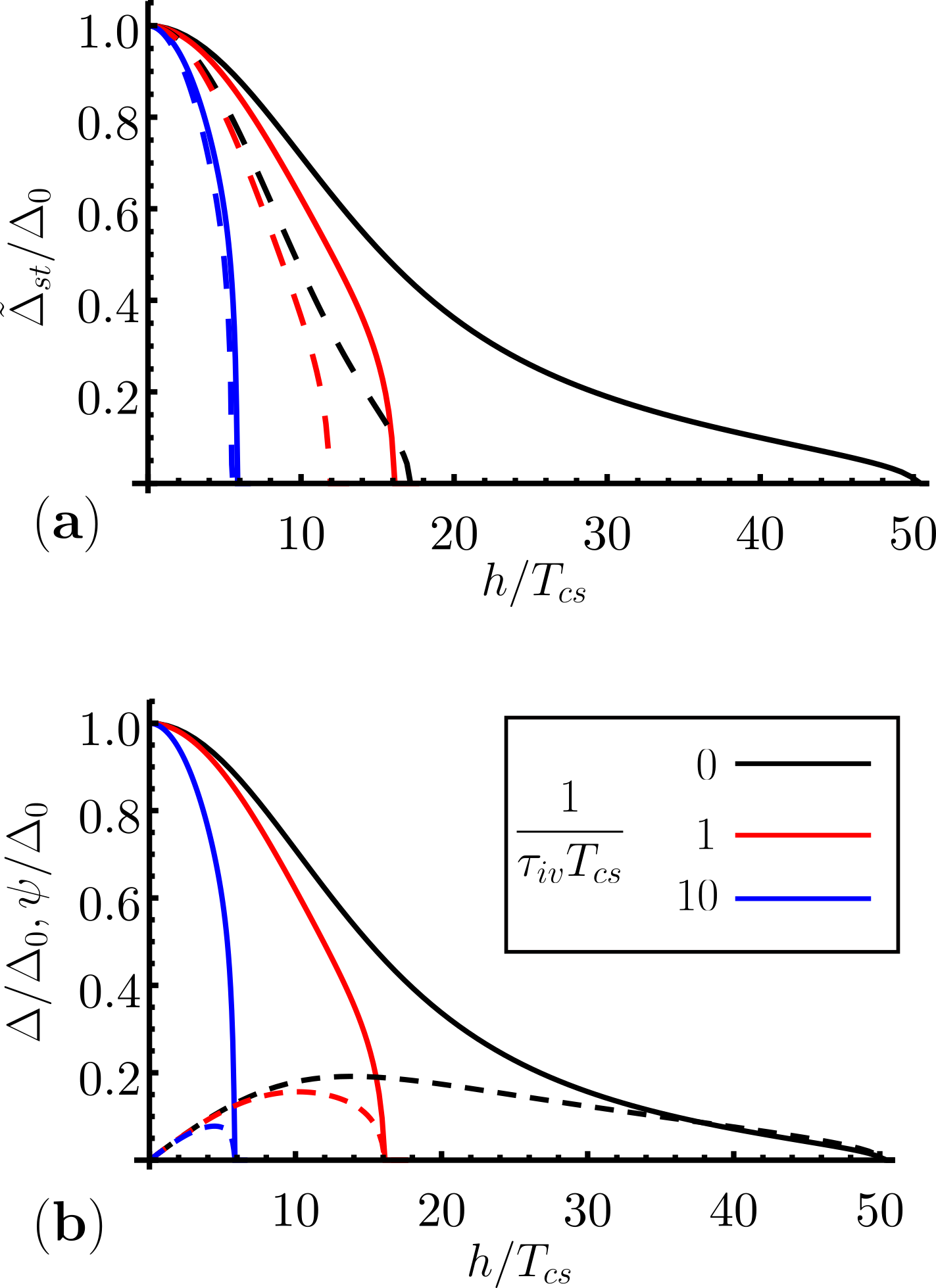}
     \caption{(a) Renormalized order parameter $\tilde{\Delta}_{st}$ as a function of Zeeman field, for different strengths of intervalley disorder. Dashed and full lines correspond to scenarios without and with triplet pairing ($T_{ct}=0.05 T_{cs}$), respectively. (b) For the scenario with triplet pairing, we plot separately the singlet and triplet order parameter, represented by the full and dashed lines, respectively. In all plots $\Delta_{so}=20 T_{cs}$ and the temperature is set to $T=0.1 T_{cs}$, where $T_{cs}$ is the singlet critical temperature. 
     \label{FigDelta}
     }
\end{figure}

In Fig.~\ref{FigDelta}, we show results of the self-consistent calculation of the order parameters as a function of magnetic field for different values of intervalley disorder. First, using Eqs.~\eqref{eq-AG} and \eqref{eq-Deltast}, we determine the effective order parameter $\tilde{\Delta}_{st}$, shown in panel (a). Then, using the first and second line of Eq.~\eqref{eq-self}, we can separately extract the singlet and triplet order parameter, respectively, shown in panel (b). As seen from the plots, the presence of weak triplet pairing makes superconductivity significantly more robust to the effect of in-plane fields. Moreover, due to the singlet-triplet coupling by the Zeeman field, the singlet order parameter can survive far higher fields in the presence of triplets. The presence of intervalley scattering significantly reduces the upper critical field (see also Sec.~\ref{Sec4b}), and suppresses the triplet order parameter. 

At $T=0$ and in the gapped phase ($\tilde{\alpha}<\tilde{\Delta}_{st}$), from Eq.~\eqref{eq-Deltast} we can obtain an analytical expression determining the renormalized gap as
\begin{multline}
\frac{\rho^2\left(\ln\frac{\tilde\Delta_{st}}{\Delta_0}+\frac{\pi\tilde\alpha}{4\tilde\Delta_{st}}\right)\left(\ln\frac{\tilde\Delta_{st}}{\psi_0}+\frac{\pi\tilde\alpha}{4\tilde\Delta_{st}}\right)}{\Delta_{so}^2\left(\ln\frac{\tilde\Delta_{st}}{\Delta_0}+\frac{\pi\tilde\alpha}{4\tilde\Delta_{st}}\right)+h^2\left(\ln\frac{\tilde\Delta_{st}}{\psi_0}+\frac{\pi\tilde\alpha}{4\tilde\Delta_{st}}\right)}
=\\
\ln\frac{\tilde\Delta_{st}}{2\rho}+\frac{\pi\tilde\alpha}{4\tilde\Delta_{st}}.
\label{eq-zeroT}
\end{multline}
In the absence of intervalley scattering, this becomes 
\begin{eqnarray}
\tilde\Delta_{st}
&=&\Delta_0\exp\left[-\frac{h^2\ln\frac{\Delta_0}{\psi_0}\ln\frac{2\rho}{\Delta_0}}{\rho^2\ln\frac{2\rho}{\Delta_0}+\Delta_{so}^2\ln\frac{\Delta_0}{\psi_0}}\right].
\end{eqnarray}
This equation illustrates that the order parameter cannot be fully suppressed by applying the Zeeman field in the absence of intervalley disorder. This means that the upper critical field $h_{c2}$ diverges at zero temperature, as will be discussed in Sec.~\ref{Sec4b}.

\subsection{Upper critical field \label{Sec4c}}
To calculate the upper critical field $h_{c2}$, we  use the fact that the phase transition to the normal state is of the second order \cite{ilic2017enhancement, sohn2018unusual}, so that the order parameters vanish at the transition. Then, we may keep only the terms up to the linear order in $\Delta$ and $\psi$ in Eqs.~\eqref{eq-AG} and  \eqref{eq-self}.  Solving Eq.~\eqref{eq-AG} yields $u=(\omega+\tilde{\alpha})/\tilde{\Delta}_{st}$, so that $1/\sqrt{1+u^2}\approx \tilde{\Delta}_{st}/(\omega+\tilde{\alpha})$. Substituting this into Eq.~\eqref{eq-self}, and using that the coupling constants can be expressed in terms of singlet and triplet critical temperatures, $g_s^{-1}=2\pi T \sum_{\omega_n}1/\omega_n-\ln T_{cs}/T$ and $g_t^{-1}=2\pi T \sum_{\omega_n}1/\omega_n-\ln T_{ct}/T$, we can write the linearized gap equation as
\begin{equation}
\det[\hat{T}+\hat{\mathcal{S}}]=0,
\end{equation}
 where
\begin{equation}
\hat{T}=\begin{bmatrix}
\ln\frac{T}{T_{cs}} & 0 \\
0 & \ln\frac{T}{T_{ct}}
\end{bmatrix}, 
\label{eq-matT}
\end{equation}
and 
\begin{multline}
\hat{\mathcal{S}}=\Psi \bigg(\frac{i\rho_{c2}}{2\pi T}\bigg) \frac{1}{\rho_{c2}^2}
\begin{bmatrix}
h_{c2}^2 & \Delta_{so}h_{c2} \\
\Delta_{so}h_{c2} & \Delta_{so}^2
\end{bmatrix} \\
+
\Psi\bigg(\frac{\tilde{\alpha}(h_{c2})}{2\pi T}\bigg)\frac1{\rho_{c2}^2}
\begin{bmatrix}
\Delta_{so}^2 & -\Delta_{so}h_{c2} \\
-\Delta_{so}h_{c2} & h_{c2}^2
\label{eq-hc2}
\end{bmatrix}. 
\end{multline}
Here  $\Psi(x)=\text{Re}[\psi(\frac{1}{2})-\psi(\frac{1}{2}+x)]$, where $\psi(x)$ is the digamma function, and we introduced $\rho_{c2}=\sqrt{\Delta_{so}^2+h_{c2}^2}$.
Note that a more general expression for $h_{c2}$, valid at arbitrary $\Delta_{so}$, has been presented in Ref.~\onlinecite{mockli2020ising} (see also Appendix \ref{App2}).

\begin{figure}
    \includegraphics[width=0.35\textwidth]{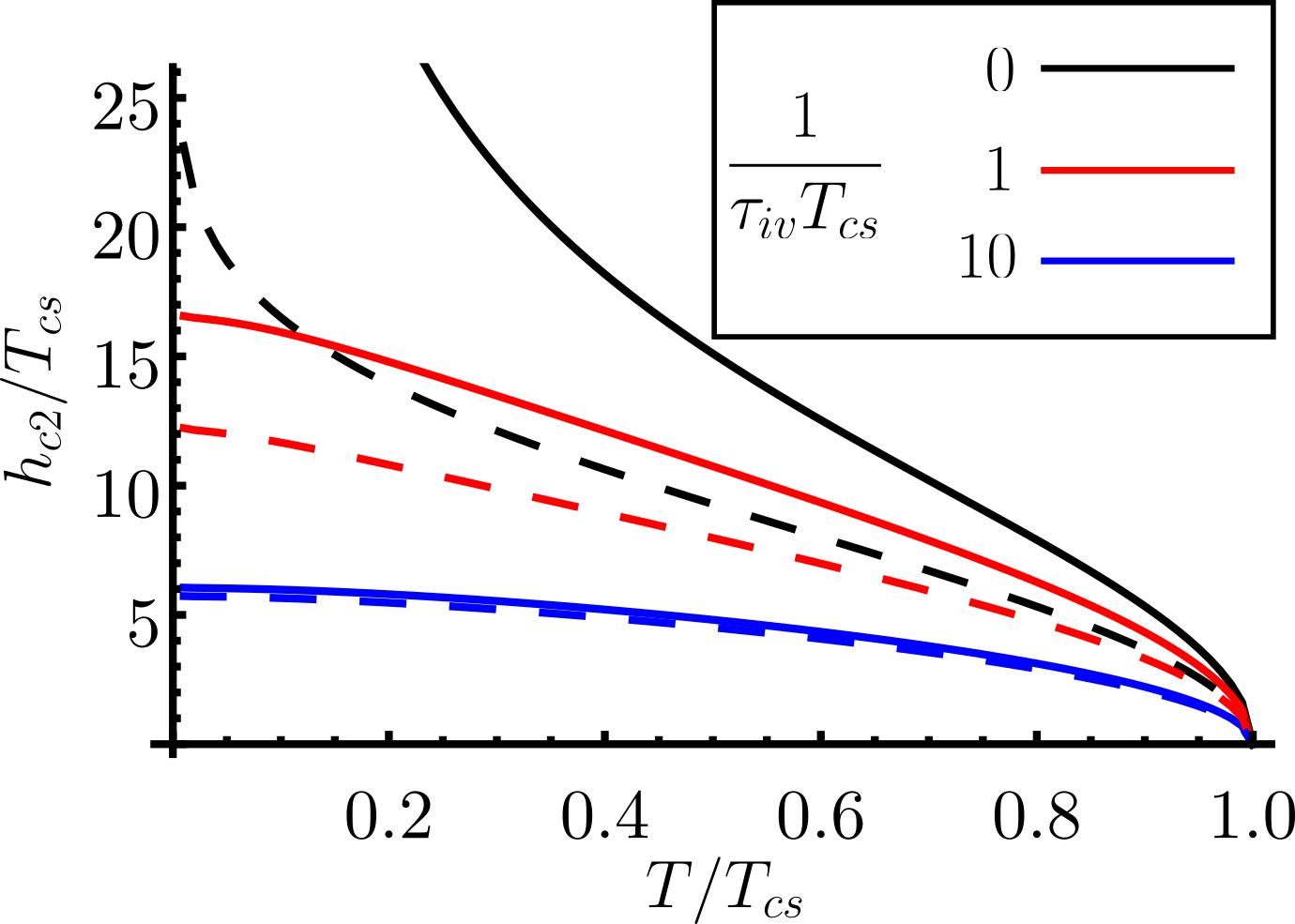}
     \caption{Upper critical field as a function of temperature, for different values of intervalley scattering. Spin-orbit coupling is set to $\Delta_{so}=20 T_{cs}$. Dashed and full lines correspond to scenarios without and with triplet pairing ($T_{ct}=0.05 T_{cs}$), respectively. \label{Fighc2} }
\end{figure}

In the absence of intervalley disorder, only the first line is finite in Eq.~\eqref{eq-hc2}. In that case $h_{c2}$ logarithmically diverges at zero temperature \cite{ilic2017enhancement} in the absence of triplet pairing. The triplet pairing shifts this divergence to $T_{ct}$, and the critical field is infinite at all $T<T_{ct}$. At finite intervalley disorder, the depairing term in the second line of Eq.~\eqref{eq-hc2} cuts-off the divergence. If the intervalley scattering is strong, $\tau_{iv}^{-1}\gg \Delta_0$, the triplets are fully  suppressed. Moreover, in this regime $\Delta_{so}\gg h_{c2}$, so only the second line in Eq.~\eqref{eq-hc2} is relevant, and we obtain the standard AG depairing equation for the critical field 
\begin{equation}
\ln \frac{T}{T_{cs}}=\Psi \bigg(\frac{\alpha(h_{c2})}{2\pi T}\bigg).
\end{equation}
In Fig.~\ref{Fighc2}, we illustrate the behavior of $h_{c2}$ as a function of temperature at different values of intervalley scattering. We see that weak triplet pairing significantly increases the $h_{c2}$ in the absence of intervalley disorder. Including such disorder quickly suppresses the effect of triplets.

\subsection{Density of states \label{Sec4d}}
The density of states is given as
\begin{equation}
\nu(\epsilon)=\nu_{0}\text{Re}\left[\frac{u}{\sqrt{1+u^2}}\right]_{i\omega_n\to \epsilon+i0^+}, 
\label{eq-dos}
\end{equation}
where $u$ is obtained as a solution of Eq.~\eqref{eq-AG}. The gap edge in the DoS is given by the standard expression~\cite{abrikosov1960contribution, maki2018gapless}: $E_g^{2/3}=\tilde{\Delta}_{st}^{2/3}-\tilde{\alpha}^{2/3}$. Note that the DoS calculated this way is correct only at low energies, and the high-energy features of the spectrum, the  \lq\lq mirage\rq\rq\ gaps, are not captured. We discuss these features in more detail in Sec.~\ref{Sec5}.

In the absence of intervalley scattering, the DoS is given by sharp BCS peaks at the renormalized gap $\tilde{\Delta}_{st}$
\begin{equation}
\nu(\epsilon)=\nu_{0}\frac{|\epsilon|}{\sqrt{\epsilon^2-\tilde\Delta_{st}^2}}\theta(|\epsilon|-\tilde\Delta_{st}).
\label{Eq:DosClean}
\end{equation}
In contrast to conventional superconductors, which exhibit spin splitting of the coherence peaks when the Zeeman field is applied, clean Ising superconductors will have a single coherence peak even at high fields. The only effect of increasing the field is the reduction of the gap in the quasiparticle spectrum.

\begin{figure}
    \includegraphics[width=0.40\textwidth]{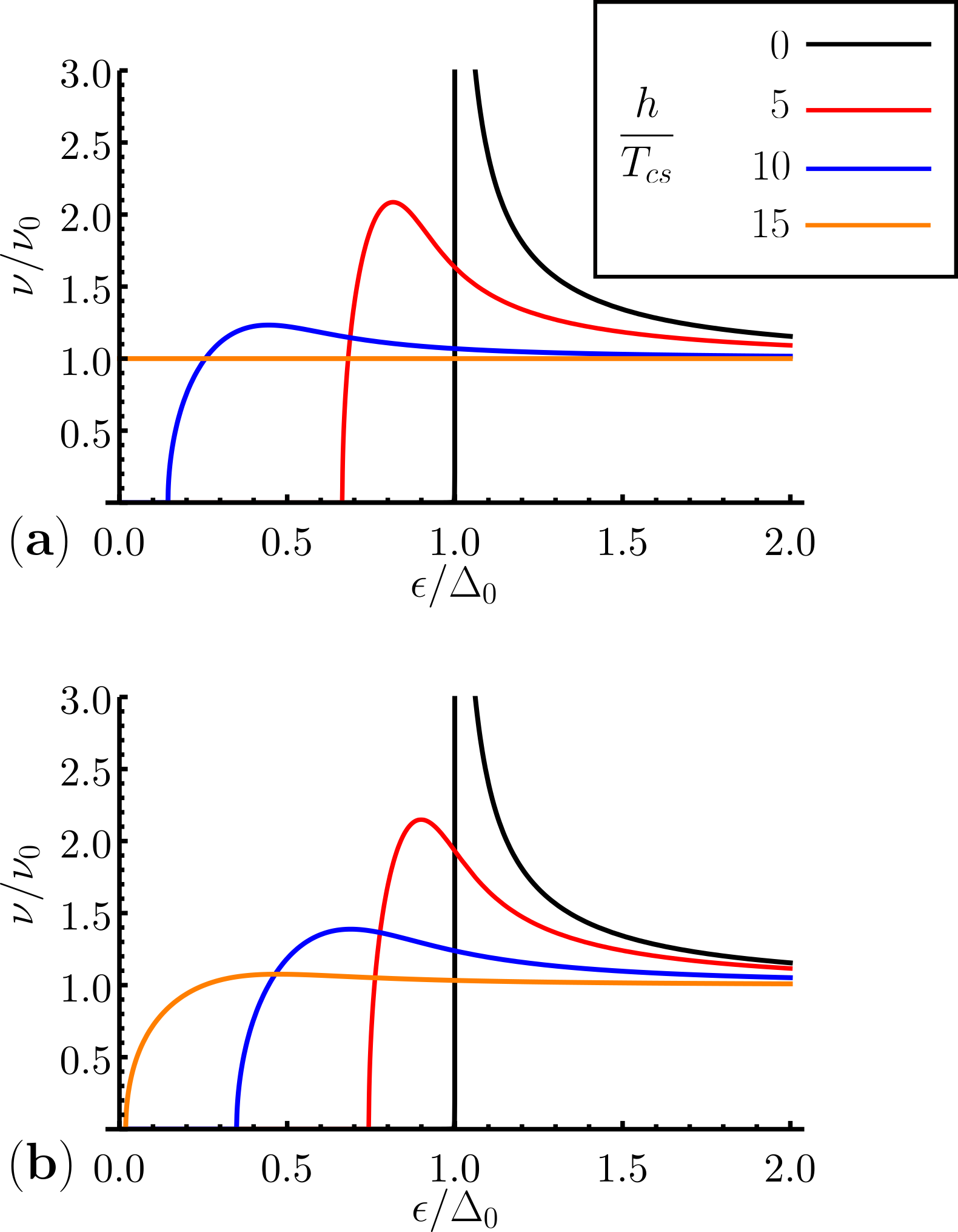}
     \caption{Density of states of an Ising superconductor for different values of the in-plane field: (a) without triplet pairing, and (b) with triplet pairing ($T_{ct}=0.05 T_{cs}$). We set $\Delta_{so}=20 T_{cs}$, $T=0.1 T_{cs}$ and $\tau_{iv}^{-1}=T_{cs}$. DoS is plotted using self-consistent order parameters $\Delta$ and $\psi$ at each value of the Zeeman field (see Fig. \ref{FigDelta} ).\label{FigDoS}}
\end{figure}

Adding intervalley scattering introduces a depairing mechanism and the DoS becomes smeared \cite{haim2020signatures}. In Fig.~\ref{FigDoS}, we plot the DoS for different values of in-plane field, using the self-consistent order parameters from Fig.~\ref{FigDelta}. As seen from the plots, the gap in the DoS is significantly more robust to the fields in the presence of triplets, due to the increased robustness of the order parameters $\Delta$ and $\psi$ (see Fig.~\ref{FigDelta}).

\section{\lq\lq Mirage\rq\rq\ gaps at high energies \label{Sec5}}

\begin{figure}[h!]
    \includegraphics[width=0.35\textwidth]{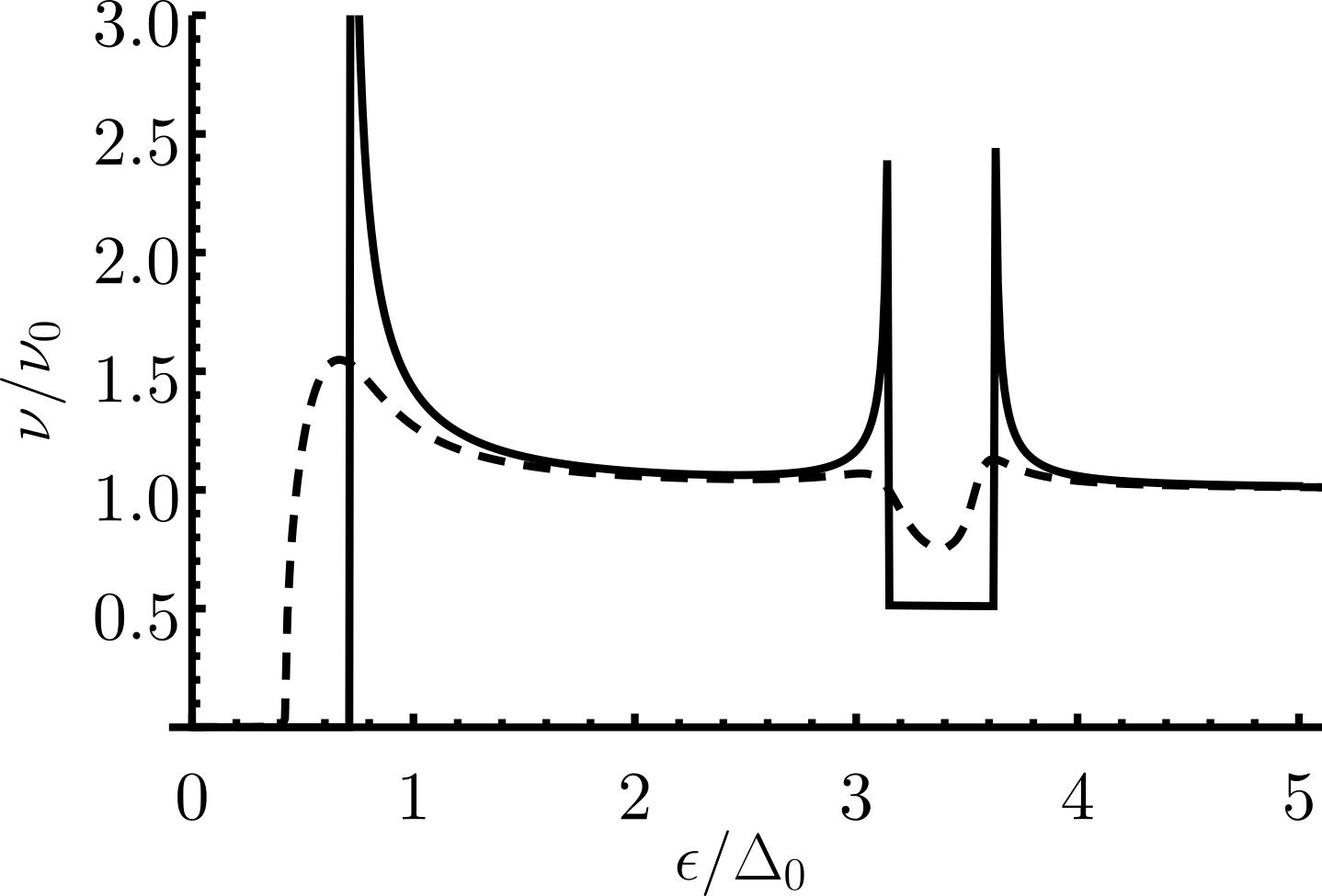}
     \caption{Mirage gap of the Ising superconductor with triplet pairing in the absence of disorder (full line) and in the presence of weak disorder (dashed line, $\tau_{iv}^{-1}=0.5 T_{cs}$). The parameters used in the plots are $\Delta_{so}=5 T_{cs}$, $h=3 T_{cs}$, $T=0.1 T_{cs}$ and $T_{ct}=0.05 T_{cs}$.  \label{figMirage}} 
\end{figure}

Mirage gaps are high-energy half-gap features in the DoS of Ising superconductors introduced in Ref.~\onlinecite{tang2021magnetic} (see Fig.~\ref{figMirage}), which appear as a consequence of equal-spin triplet correlations of the form $\sim | \downarrow \downarrow \rangle +|\uparrow \uparrow \rangle$.  In pure singlet-pairing Ising superconductors this type of correlations is enabled by the combination of  ISOC and  Zeeman field. Finite equal-spin triplet additionally contributes as a source of such correlations \cite{kuzmanovic2022tunneling, patil2023spectral}.  

Formally, the correlations responsible for the mirage gaps are captured by the term $\sim s_y \tau_x$ in the quasiclassical Green's function $g_\eta$ [see Eq.~\eqref{eq-5}]. Since these correlations are odd in valley index $\eta$, they are sensitive to intervalley disorder. This is illustrated in Fig.~\ref{figMirage}, where we show how the mirage gaps are  suppressed by weak intervalley scattering.

At large ISOC, the mirage gaps appear at energies  $\epsilon\approx \pm \rho$. Assuming $\Delta_{so}, \epsilon \gg \Delta_0, \tau_{iv}^{-1},h$, from Eqs.~\eqref{eq-5} and \eqref{eq-6} we find an approximate expression for the DoS as
\begin{equation}
\nu\approx \frac{\nu_0}{2} \text{Re} \left[
1+\frac{|\tilde{\Sigma}|}{\sqrt{\tilde{\Sigma}^2-4P^2}}
\right] \bigg|_{i\omega_n\to \epsilon+i 0^+},
\end{equation}
where $\tilde{\Sigma}=\omega_n(\omega_n+\tau_{iv}^{-1})+\rho^2+\Delta^2+\psi^2$ and $P=h\Delta-\Delta_{so}\psi$. 
 In the absence of intervalley scattering, the width of the mirage gaps can be approximated as
\begin{equation}
W\approx \frac{2}{\rho}(h\Delta-\Delta_{so}\psi).
\end{equation}
The ratio of the mirage gap and the main gap is then $W/(2\tilde\Delta_{st})\approx h/\Delta_{so}-\rho\psi/(\Delta_{so}\tilde\Delta_{st})$. Therefore, the locking of the relative phase of singlet and triplet order parameters ($-\pi/2$) is such that it increases the width of the main gap, but reduces the width of the mirage gap.

\begin{figure}
    \includegraphics[width=0.5\textwidth]{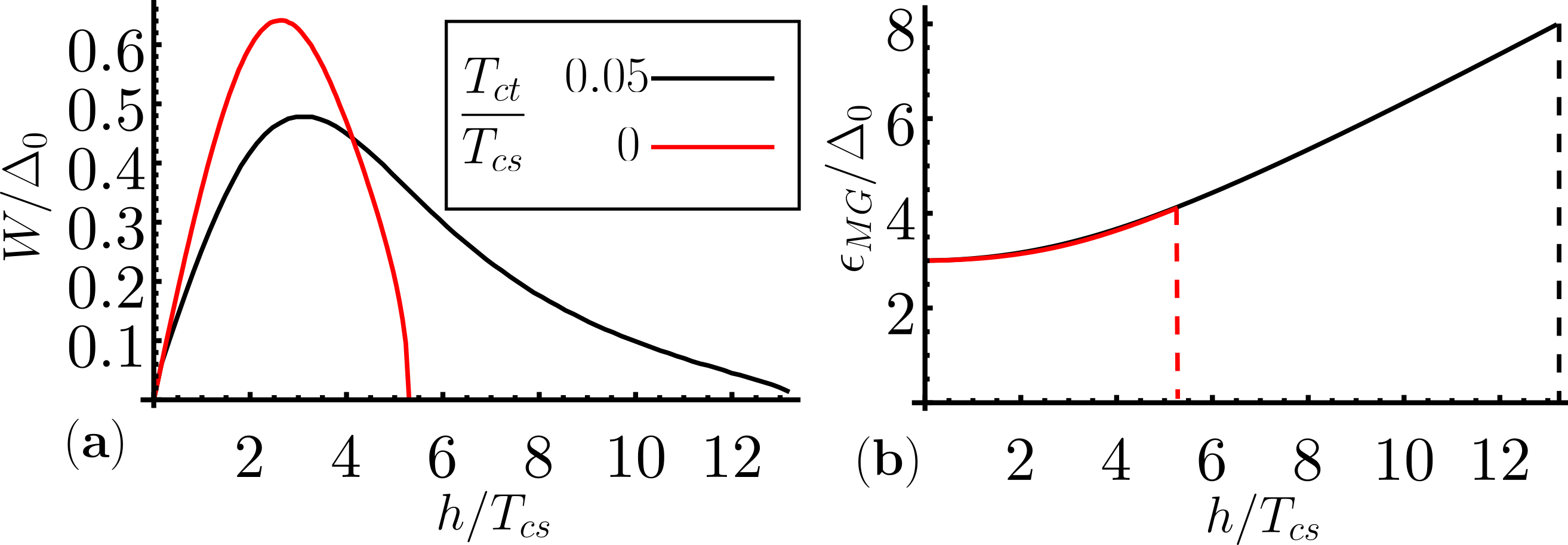}
     \caption{(a) Width of the mirage gaps, and (b) their position, as a function of the in-plane magnetic field in the absence of intervalley disorder. We consider the scenario with and without triplet pairing ($T_{ct}=0.05 T_{cs}$). The parameters used in the plots are $\Delta_{so}=5 T_{cs}$, $T=0.1 T_{cs}$. 
    Vertical dashed lines in panel (b) indicate the upper critical fields.
     \label{figMirageW}
     }
\end{figure}

In the absence of intervalley disorder,  the mirage gaps can be straightforwardly calculated at arbitrary ISOC using analytical expressions for the DoS and the self-consistency condition first presented in  Ref.~\onlinecite{kuzmanovic2022tunneling} (see also Appendix \ref{App1} and Ref.~\onlinecite{patil2023spectral}).
In Fig.~\ref{figMirageW}(a), we plot the width of the mirage gap as a function of magnetic field with and without triplet pairing. At low fields, we see that the presence of triplet pairing reduces the width of mirage gaps.  Moreover, in the presence of triplets the mirage gaps are able to persist up to much higher fields, due to the increased $h_{c2}$. These findings are consistent with the results of Ref.~\onlinecite{patil2023spectral}. The position of mirage gaps $\epsilon_{MG}$, illustrated in Fig.~\ref{figMirageW}(b), is not significantly influenced by the presence of triplet pairing.

\section{Conclusions \label{Sec7}}
In this work, we showed that the complex interplay of ISOC, Zeeman field, disorder, and mixed singlet-triplet pairing in Ising superconductors can be captured by relatively simple equations resembling the well-known Abrikosov-Gor'kov theory in the realistic regime of strong ISOC. We use our results to calculate the bulk properties of the superconductor such as the upper critical field and the DoS, which can be directly useful for the recent experiments measuring these quantities. Moreover, our equations could be useful for future studies of hybrid structures of Ising superconductors, such as Josephson junctions or superconductor/ferromagnet hybrids, which have recently been experimentally realized in van der Waals heterostructures \cite{hamill2021two, idzuchi2021unconventional, kang2021giant}. 

\section*{Acknowledgements}
We thank M. Khodas and M. Haim for interesting discussions at various stages of the project. SI acknowledges funding from the Laboratoire d’excellence LANEF in Grenoble (Grant No. ANR-10-LABX-0051) and the European Union's Horizon 2020
Research and Innovation Framework Programme under
Grant No. 800923 (SUPERTED). JM acknowledges funding from the French Agence Nationale de la Recherche through Grant No. ANR-21-CE30-0035 (TRIPRES). 

\appendix
\section{Analytical expressions in the absence of intervalley disorder at arbitrary ISOC strength \label{App1}}
 In the absence of intervalley scattering, it is possible to obtain an analytical solution for the quasiclassical Green's function from Eqs.~\eqref{eq-5} and \eqref{eq-6}, without making additional assumptions about the magnitude of ISOC (see also the Supplemental Material of Ref.~\onlinecite{kuzmanovic2022tunneling}). From there, we obtain for the DoS
 \begin{widetext}
\begin{equation}
\nu(\epsilon)
=2\nu_{0}\text{Re}\left[\frac{\omega_n{\rm\, sign}(\Sigma)}{X}\left(1+\frac{|\Sigma|}{\sqrt{\Sigma^2-4P^2}}\right)\right]_{i\omega_n\to\epsilon+i0^+},\label{eq-dosf}
\end{equation}
and the coupled self-consistency conditions are
\begin{eqnarray}
\Delta
&=&2\pi T g_s \sum_{\omega_n>0} \frac1{X}\left[\Delta\left(1+\frac{\Sigma-2h^2}{\sqrt{\Sigma^2-4P^2}}\right)+\psi\frac{2 h \Delta_{so}}{\sqrt{\Sigma^2-4P^2}}\right],\label{eq-s}\\
\psi
&=&2\pi T g_t \sum_{\omega_n>0} \frac1{X}\left[\psi\left(1+\frac{\Sigma-2\Delta_{so}^2}{\sqrt{\Sigma^2-4P^2}}\right)+\Delta\frac{2h\Delta_{so}}{\sqrt{\Sigma^2-4P^2}}\right].\label{eq-t}
\end{eqnarray}
\end{widetext}
We introduced the notation $\Sigma=\omega_n^2+\rho^2+\Delta^2+\psi^2$, $P=h\Delta-\Delta_{so}\psi$, and $X=\sqrt2\left[\Sigma-2\rho^2 +{\rm\, sign}(\Sigma)\sqrt{\Sigma^2-4P^2}\right]^{1/2}.$

\section{Strongly disordered limit $\tau_{iv}^{-1} \gtrsim \Delta_{so}\gg \Delta$ \label{App3}}
In the main text, we focused on the realistic regime of strong ISOC, $\Delta_{so}\gg \tau_{iv}^{-1},\Delta$. Here, we will consider the opposite regime $\tau_{iv}^{-1} \gtrsim \Delta_{so}\gg \Delta$. 

For strong-enough intervalley scattering,  $\tau_{iv}^{-1}\gg \Delta$, triplet pairing is suppressed, as shown in the main text. In this regime, the general equations \eqref{eq-6} and \eqref{eq-7} reduce to
\begin{equation}
\frac{\omega_n\pm ih}{\Delta}=u_{\pm}+\frac{\Delta_{so}^2\tau_{iv}}{2\Delta}\frac{u_\pm-u_\mp}{\sqrt{1+u_{\mp}^2}}, 
\label{eqa1}
\end{equation}
 where $2 x=\sum_\pm u_\pm (1+u_\pm^2)^{-1/2}$ and $2y= x^{-1}\sum_\pm(1+u_\pm^2)^{-1/2}$.
This is equivalent to the result of Maki and Tsuneto \cite{maki1964pauli} for 2D superconductors with spin-orbit impurities. Namely, the effect of ISOC is fully captured by an effective spin-orbit scattering rate $\Delta_{so}^2\tau_{iv}$. Note that a similar effective scattering rate also appears in the studies of weak localization of TMD materials in the relevant regime \cite{ilic2019weak}. 

 As in Ref.~\onlinecite{maki1964pauli}, the DoS is given as
\begin{equation}
\nu(\epsilon)=\frac{\nu_{0}}{2}\sum_\pm \text{Re}\left[\frac{u_\pm}{\sqrt{1+u_\pm^2}}\right]_{i\omega_n\to\epsilon+i0^+},
\label{eqa2}
\end{equation}
 the self-consistency condition is
\begin{equation}
\Delta=\pi T g_s\sum_{\omega_n>0,\pm} \frac1{\sqrt{1+u_\pm^2}},
\label{eqa3}
\end{equation}
 and the linearized gap equation is 
\begin{equation}
\ln\frac{T}{T_{cs}}=\textcolor{black}{2\pi T} \sum_{\omega_n}\bigg[\frac{1}{\omega_n}-\frac{\omega_n+\Delta_{so}^2\tau_{iv}}{h^2+\omega_n (\omega_n+\Delta_{so}^2\tau_{iv})}\bigg].
\label{eqa4}
\end{equation}

Note that Eq.~\eqref{eqa1} reduces to the Abrikosov-Gor'kov equation if ISOC is strong enough, $\Delta_{so}^2\tau_{iv}\gg\Delta_0$, with a depairing parameter $\alpha=h^2/(\Delta_{so}^2 \tau_{iv})$ and the gap parameter $\Delta$. Therefore, there is an overlap with the regime $\Delta_{so}\gg \tau_{iv}^{-1}\gg \Delta_0$ presented in Eqs.~\eqref{eq-17} and \eqref{eq-18} of the main text.

On the other hand, for weak ISOC, $\Delta_{so}^2\tau_{iv}\ll \Delta_0$, we recover the standard equations for the paramagnetically limited superconductor. Here,  the electrons are so frequently scattered between the valleys that they no longer \lq\lq feel\rq\rq\ the ISOC.  In this regime the DoS is spin-split 
\begin{equation}
\nu(\epsilon)=\frac{\nu_0}{2}\sum_\pm \text{Re} \frac{\epsilon\pm h}{\sqrt{(\epsilon\pm h)^2-\Delta^2}},
\end{equation}
and the linearized gap equation is:
\begin{equation}
\ln\frac{T_{cs}}{T}=\text{Re}\bigg[\psi\bigg(\frac{1}{2}+\frac{i h}{2\pi T}\bigg)-\psi\bigg(\frac{1}{2}\bigg)\bigg].
\end{equation}

\section{Upper critical field at arbitrary ISOC strength \label{App2}}
In Sec.~\ref{Sec4b} we presented an analytical result determining  the upper critical field at strong ISOC $\Delta_{so}\gg \Delta, \tau_{iv}^{-1}$. Here, we will derive a more general expression valid for any ISOC strength. 

In the vicinity of the transition to the normal state, we linearize Eq.~\eqref{eq-6} in order parameters and solve it to determine $y$ by taking $x\approx 1$. Then, we can obtain the coupled linear gap equations from Eq.~\eqref{eq-selfcons}
\begin{eqnarray}
\Delta&=&2\pi Tg_s\sum_{\omega_n>0}\bigg[\Delta\frac{\omega_n(\omega_n+\frac1{\tau_{iv}})+\Delta_{so}^2}{\mathcal{A}_n}+\psi\frac{\Delta_{so}h_{c2}}{\mathcal{A}_n}\bigg], \nonumber \\
\psi&=&2\pi Tg_t\sum_{\omega_n>0}\bigg[\Delta\frac{\Delta_{so}h_{c2}}{\mathcal{A}_n}+\psi\frac{\omega_n^2+h_{c2}^2}{\mathcal{A}_n}\bigg].
\end{eqnarray}
Here, we introduced $\mathcal{A}_n=(\omega_n+\frac1{\tau_{iv}})(\omega_n^2+\rho_{c2}^2)-\frac1{\tau_{iv}}\Delta_{so}^2$. Then, the linear gap equation can be written in the form $\text{det}[\hat{T}+\hat{\mathcal{S}}]=0$, where $\hat{T}$ was introduced in Eq.~\eqref{eq-matT}, and 
 \begin{equation}
 \hat{\mathcal{S}}= \begin{bmatrix}
\mathcal{S}_{s} & \mathcal{S}_{st} \\
\mathcal{S}_{st} & \mathcal{S}_{t}
\end{bmatrix},
\label{eq-S1}
 \end{equation}
 where
\begin{align}
&\mathcal{S}_s=2\pi T\sum_{\omega_n>0}\bigg[\frac{1}{\omega_n}-\frac{\omega_n(\omega_n+\frac1{\tau_{iv}})+\Delta_{so}^2}{\mathcal{A}_n} \bigg],  \nonumber \\
& \mathcal{S}_t=2\pi T\sum_{\omega_n>0}\bigg[\frac{1}{\omega_n}-\frac{\omega_n^2+h_{c2}^2}{\mathcal{A}_n}\bigg], \nonumber\\
&\mathcal{S}_{st}=2\pi T\sum_{\omega_n>0}\frac{\Delta_{so}h_{c2}}{\mathcal{A}_n}.
\label{eq-S2}
\end{align}
Similar expressions were previously reported in Ref.~\onlinecite{mockli2020ising}. In the case of singlet-pairing only, the linear gap equation reduces to  $\ln(T_{cs}/T)=\mathcal{S}_s$ \cite{ilic2017enhancement}. Taking the limit $\Delta_{so}\gg \Delta, \tau_{iv}^{-1}$, Eqs.~\eqref{eq-S1} and \eqref{eq-S2} reduce to Eq.~\eqref{eq-hc2} from the main text.

\end{document}